\documentclass[aps,prd,twocolumn,showpacs,groupedaddress,nofootinbib,letterpaper]{revtex4}

\usepackage{amsmath}
\usepackage{amsfonts}
\usepackage{amssymb}

\usepackage{hyperref}

\def\bea{\begin{eqnarray}}
\def\eea{\end{eqnarray}}
\def\beq{\begin{equation}}
\def\eeq{\end{equation}}
\def\nn{\nonumber}

\begin{document}

\title{Hamiltonian structure of Ho\v{r}ava gravity}
\author{William Donnelly}
\email{wdonnell@umd.edu}
\author{Ted Jacobson}
\email{jacobson@umd.edu}
\affiliation{
Center for Fundamental Physics\\
Department of Physics \\
University of Maryland at College Park \\
College Park, Maryland, 20742-4111 USA
}
\begin{abstract}
The Hamiltonian formulation of Ho\v{r}ava gravity is derived.
In a closed universe the Hamiltonian is a sum of 
generators of gauge symmetries, the foliation-preserving diffeomorphisms,
and vanishes on shell.
The scalar constraint is second class, except for a global, first-class
part that generates time reparametrizations.
A reduced phase space formulation is given in which 
the local part of the scalar constraint is solved formally for the lapse as a function
of the 3 metric and its conjugate momentum. In the infrared limit
the scalar constraint is linear in the square root of the lapse. 
For asymptotically flat boundary conditions the Hamiltonian is 
a sum of bulk constraints plus a boundary term that gives the total energy.
This energy expression is identical to the one for Einstein-aether theory
which, for static spherically symmetric solutions, is the 
usual Arnowitt-Deser-Misner energy of general relativity with a rescaled Newton constant.
\end{abstract}

\pacs{04.50.Kd, 04.20.Fy}

\maketitle

\section{Introduction}
Ho\v{r}ava \cite{Horava2009} has proposed a theory of gravity that is closely
related to general relativity (GR), but is power-counting renormalizable
and possesses a preferred spacelike foliation of spacetime that breaks the spacetime
diffeomorphism symmetry down to time-dependent
3-dimensional diffeomorphism symmetry and a reparametrization
of global time.
Several variants of this theory have been considered (see Ref.~\cite{Sotiriou2010} for a review). 
In this paper we study
the ``consistent extension'' of Blas, Pujol\`as and Sibiryakov \cite{Blas2009b}.
This theory is just the ``nonprojectable" version of Ho\v{r}ava's original proposal with the inclusion of terms involving $(\ln N)_{,i}$, the spatial gradient of the log of the lapse function $N$,
which are compatible with the symmetry but were not explicitly mentioned in 
Ref.~\cite{Horava2009}. 
This variant of the theory is free of known pathologies (instabilities, overconstrained evolution, or strong coupling at low energies) which afflict some other variants. 
We will refer to this theory simply as Ho\v{r}ava gravity.

In this paper, we consider the Hamiltonian formulation of Ho\v{r}ava gravity.
While this was already considered in Ref.~\cite{Kluson2010b} (see also comments in Ref.~\cite{Blas2009b}), there are several reasons to revisit the analysis.
In Ref.~\cite{Kluson2010b}, it was argued that, unlike other theories with time-reparametrization symmetry, Ho\v{r}ava gravity has a nonvanishing Hamiltonian.
A vanishing Hamiltonian is one aspect of the so-called ``problem of time'' in quantum gravity: quantization leads to a theory in which there is no unitary evolution with respect to an external time \cite{Isham1992}.
Here we find that although the Hamiltonian density is a sum of second-class constraints, the total Hamiltonian is a sum of first-class constraints, one of which was overlooked previously.
Once this constraint is recognized, we see that the Hamiltonian for a closed space is indeed a sum of constraints, and therefore a global version of the problem of time persists.

We also consider the Hamiltonian formulation with asymptotically flat boundary conditions.
In Ref.~\cite{Jacobson2010} it was shown that in the IR limit 
Ho\v{r}ava gravity is closely related to Einstein-aether theory with a hypersurface-orthogonal aether field, in the sense that every hypersurface-orthogonal aether solution is also a Ho\v{r}ava solution.\footnote{In fact there can be a global topological obstruction to this equivalence. 
If $\Sigma$ has a nontrivial first homotopy group then it is possible to have an aether field that is locally hypersurface orthogonal but is not normal to any global foliation.
A simple example is provided by the ``tilted aether" Bianchi type I cosmologies discussed in Ref.~\cite{Carruthers2010}, if a homogeneous aether field on a homogeneous spacetime with topology $R^3\times S^1$ is tilted along the $S^1$ direction.}
We show that the energy of asymptotically flat solutions of Ho\v{r}ava gravity can be expressed as a surface integral at spatial infinity that agrees with the expression for energy in Einstein-aether theory.

\section{Ho\v{r}ava gravity} \label{horava}

In a covariant formulation~\cite{Blas2009a}, 
the dynamical objects of Ho\v{r}ava gravity are a 4 metric and a preferred foliation.
In the Hamiltonian formulation one specifies an additional foliation, which need not coincide with the preferred foliation.
However, it will be useful to use the preferred foliation in the Hamiltonian formulation, since then the Lagrangian depends only on first time derivatives of the fields.
We will therefore present the theory in coordinates $(t, x^i)$ adapted to the preferred foliation. This is the original formulation of Ho\v{r}ava\cite{Horava2009}.\footnote{It can happen 
that such coordinates do not cover the complete manifold as, 
for example, in black hole formation by collapse \cite{Barausse2011}.}
In these coordinates Ho\v{r}ava gravity has the symmetry of 
foliation-preserving diffeomorphisms
\begin{equation} \label{reparameterization}
t \to t'(t), \qquad x^i \to x'{}^i(x^i, t).
\end{equation}
The metric can be expressed in Arnowitt-Deser-Misner (ADM) form in terms of the
lapse $N$, shift vector $N^i$ and three-dimensional metric $g_{ij}$,
\begin{equation}
%ds^2 = (N^2 - N_i N^i) dt^2 - 2 N_i dx^i dt - g_{ij} dx^i dx^j.
ds^2 = N^2 dt^2 - g_{ij} (dx^i + N^i dt)(dx^j + N^j dt)
\end{equation}
where letters $i,j,\ldots = 1,\ldots,3$ denote spatial indices.
The extrinsic curvature $K_{ij}$ of the constant $t$ surfaces  and its trace
$K$ are given by
\begin{equation}
K_{ij} = \frac{1}{2N} \left[ \dot g_{ij} - \nabla_i N_j - \nabla_j N_i \right], \qquad
K = g^{ij} K_{ij}
\end{equation}
where the dot denotes partial derivative with respect to $t$, 
and $\nabla_i$ is the three-dimensional covariant derivative 
compatible with the metric $g_{ij}$.
The acceleration of the congruence normal to the constant $t$ surfaces has the spatial projection
\begin{equation} \label{acceleration}
a_i \equiv \nabla_i \ln N.
\end{equation}
In terms of these variables the Lagrangian density of Ho\v{r}ava gravity is
\begin{equation} \label{action}
\mathcal{L} = \frac{1}{16\pi G_H}\sqrt{g} N \left( K_{ij} K^{ij} - \lambda K^2 - V(g_{ij},a_i) \right),
% Note: Blas et. al. define the action with a factor of 1/2, since they use units where 8 pi G = 1
\end{equation}
where $V$ is a potential that depends on the 3 metric and the acceleration.
The potential contains all terms that are spatial scalars of dimension up to six, where the spatial coordinates $x^i$ are taken to have dimension $-1$.
In the infrared (IR) limit, the potential is dominated by the lowest dimension terms
\begin{equation} \label{potential}
V(g_{ij}, a_i) =- \xi R - \alpha a_i a^i  + \ldots,
\end{equation}
where $R$ is the Ricci scalar of the 3 metric $g_{ij}$, and $\ldots$ refers to terms containing more than two spatial derivatives.
The term $\nabla \cdot a$ is also of second order in derivatives, but its contribution to the Lagrangian differs from that of $a^2$ by a total derivative since $N \nabla \cdot a = \nabla^2 N - N a^2$.
The dimensionless free parameters of the IR limit of the theory are $\alpha$, 
$\xi$ and $\lambda$. 
When $\alpha = 0$ and $\lambda = \xi = 1$, the theory reduces to general relativity.
From here on we adopt units where $16 \pi G_H = 1$. In this paper we do not consider 
matter couplings.

\section{Hamiltonian and constraints} \label{Hamiltonian}

We now consider the Hamiltonian formulation of the theory, following Ref.~\cite{Kluson2010b}.
First, since the time derivatives of $N$ and $N^i$ do not appear in the action, the corresponding conjugate momenta $p_N$ and $p_i$ vanish, i.e. we have primary constraints
\beq\label{primary}
p_N=0,\quad p_i=0.
\eeq
The momentum conjugate to $g_{ij}$ is
\begin{equation}
p^{ij} \equiv \sqrt{g} (K^{ij} - \lambda K g^{ij}),
\end{equation}
and this relationship can be inverted yielding
\begin{equation}
K^{ij} = \frac{1}{\sqrt{g}} \left( p^{ij} + \frac{\lambda}{1 - 3 \lambda} p\, g^{ij} \right),
\end{equation}
where $p \equiv p^{ij}g_{ij}$.

The Hamiltonian density $\cal H$  has the usual form 
$p^{ij} \dot q_{ij} - \mathcal{L}$ plus
the primary constraints times Lagrange multipliers.
Up to total derivatives it takes the form
\bea \label{H}
\mathcal{H} &=& N\mathcal{H}_t +N^i\mathcal{H}_i + v^i p_i + v p_N\\
\mathcal{H}_t&=&
\frac{1}{\sqrt{g}} \left( p^{ij} p_{ij} + \frac{\lambda}{1 - 3\lambda} p^2 +gV\right)\\
\mathcal{H}_i &=&  - 2 g_{ik}\nabla_j p^{jk}
% Before integration: + 2 p^{ij}\nabla_i N_j.
% After integrations: - 2 N_j \nabla_i p^{ij}
\label{H0}
\eea
For now we will assume that $\Sigma$ has no boundary, so that total derivative terms play no role, but they will be considered in Sec.~\ref{asymflat} when asymptotically flat boundary conditions are imposed.
The Hamiltonian is then given by
\begin{equation} \label{hamiltonian}
H = \int_\Sigma d^3y\, \mathcal{H}(y) \equiv \int \mathcal{H}.
\end{equation}
Here and below
we adopt a notation in which $\int$ means 
$\int_\Sigma d^3y$ and the dependence on spatial coordinates
of integration is suppressed.

The primary constraints \eqref{primary} must be preserved in time; in other words they must have a vanishing Poisson bracket with the Hamiltonian.
Among these constraints there is a particular combination 
$\int N p_N$ that generates constant rescalings of $N$. 
Since $a_i$ is invariant under such rescalings, only explicit $N$ and 
$p_N$ dependence
need be considered. Preservation of this constraint in time thus requires that
\beq
\left \{ \int N p_N, H \right \} = \int (N \mathcal{H}_t + vp_N) = 0,
\eeq
which shows that $\int N \mathcal{H}_t = 0$ when the constraints are preserved in time. 
It follows that the Hamiltonian is a sum of constraints.
Moreover, the requirement that all constraints are preserved 
in time implies that the Hamiltonian is first class; i.e., it has zero 
Poisson brackets with all constraints.

The fact that the Hamiltonian is a sum of first-class constraints is a feature of systems that have time-reparametrization symmetry.
What distinguishes Ho\v{r}ava gravity from GR is that arbitrary deformations of the constant-time surfaces cannot be regarded as pure gauge.
In this regard, Ho\v{r}ava gravity is like a partial gauge-fixing of GR in which the global time-reparametrization freedom is left unfixed.

Preservation of the primary constraints in time leads to secondary constraints
\beq\label{secondary}
C=\delta H/\delta N =0,\quad C_i=\delta H/\delta N^i =0,
\eeq
where
\bea
\label{C}
C &=& \mathcal{H}_t - \frac{1}{N} \nabla_i V^i, \\
C_i &=& \mathcal{H}_i,
\eea
and we have defined the vector density
\beq \label{Vi}
V^i(x) = \frac{\delta}{\delta a_i(x)} \int \sqrt{g} N V.
\eeq
The vector constraint $C_i$ is the same as in GR, but the scalar constraint $C$ is modified.
Note that the constraint $C$ is invariant under a constant rescaling of $N$.
The source of this rescaling freedom is the reparametrization symmetry \eqref{reparameterization}, under which the lapse transforms as $N \to N / f'(t)$.

In the IR limit,
\beq \label{VIR}
V^i = -2 \alpha \sqrt{g} \nabla^i N,
\eeq
and the constraint $C = 0$ becomes
% Note: we divided by sqrt{g} here
\beq \label{IR}
\frac{1}{g} \left(p^{ij}p_{ij} + \frac{\lambda}{1 - 3 \lambda} p^2\right) - \xi R - \alpha \frac{(\nabla N)^2}{N^2} + 2 \alpha \frac{\nabla^2 N}{N} = 0.
\eeq
There are different strategies for solving the scalar constraint.
For example, one could solve for the conformal factor of the metric while keeping the other degrees of freedom fixed as is often done in GR.
In Ho\v{r}ava gravity there is the scalar degree of freedom $N$, and it is natural to view $C$ as an equation for $N$ \cite{Kluson2010b,Blas2009b}.
Note, however, that Eq.~\eqref{IR} can determine $N$ at most up to a constant rescaling.

The constraint equation in the IR limit can be linearized by the change of variables\footnote{This was pointed out to us by M. Grillakis.} 
$N = n^2$,
resulting in the equation
\beq \label{L1}
Ln = 0
\eeq
where $L$ is the linear differential operator
\beq \label{L2}
L \equiv -4 \alpha \nabla^2 - \frac{1}{g} \left(p^{ij}p_{ij} + \frac{\lambda}{1 - 3 \lambda} p^2\right) + \xi R.
\eeq
Such an equation admits a solution if and only if the spectrum of the Schr\"odinger-like operator $L$ contains zero.
Moreover, if the
foliation by constant $t$ surfaces is to be smooth
the lapse must be positive everywhere, 
which requires that $n(x)$ is positive for all $x$.
A solution with positive $n$ exists if and only if
 zero is the least eigenvalue of $L$: this is the familiar statement that the Schr\"odinger equation admits a unique eigenstate with no nodes, and this state is a ground state \cite{Courant1989}.
Thus Eq.~\eqref{L1} contains a condition on the metric 
$g_{ij}$ and its conjugate momentum $p^{ij}$.
If this condition is met the constraint has a unique (up to rescaling) 
positive solution for $n$, and hence the lapse is determined
up to a constant scaling.

\subsection{Propagation of constraints} \label{propagation}

The constraint equations $C = 0$ and $C_i = 0$ must hold at each time.
One way to satisfy this requirement is to solve these two equations as independent constraints at each time.
For example, one could imagine solving the constraint $C$ for $N$ as a function of the metric and its conjugate momentum, up to a time-dependent prefactor.
If this could be done, the constraint would hold for all times and there would be 
no need to add an independent condition to ensure its preservation with time.
The time evolution of $N$ is governed by the Lagrange multiplier $v$,
which is entirely free to begin with, so can always be chosen so as to produce
the required time evolution of $N$. (The situation would be quite different however if
$V$ had no $N$ dependence. Then the constraint would impose a relation 
between $g_{ij}$ and $p^{ij}$ alone that is not generally consistent with their
evolution equations, except in the GR case.)

Rather than solving the constraint at all times, one can instead solve it 
at one time, and then choose the Lagrange multiplier such that 
the constraint is preserved. We now proceed to analyze the preservation
of the constraints in this way. This approach implements the general formalism
for constrained Hamiltonian systems discussed in Ref.~\cite{Henneaux1994}, 
and allows one to identify the first-class constraints without any guesswork.

\subsubsection{Propagation of diffeomorphism constraints}

Let us consider first the diffeomorphism constraint $C_i$.
Since it generates spatial diffeomorphisms, and the Hamiltonian is a
spatial scalar (a number, not a field),
one might think that the Poisson bracket
$\{C_i,H\}$ would vanish, implying that $C_i$ is constant and therefore
the constraint $C_i=0$ is preserved in time.
There is a catch in this reasoning however, since $C_i$ only generates
diffeomorphisms of $g_{ij}$ and $p^{ij}$, whereas the Hamiltonian
also depends on $N$ and $N^i$, so in fact $\{C_i,H\}\ne0$.
The impact of the $N^i$ dependence is transparent,
since $N^i$ only enters $H$ linearly in the combination $N^iC_i$,
and the Poisson bracket algebra of diffeomorphism generators closes,
\beq
\{C_i(x),C_j(y)\}=C_i(y)\delta_{,j}(x,y)+C_j(x)\delta_{,i}(x,y).
\eeq
This $N^i$ dependence therefore produces a contribution to the time derivative
of $C_i(x)$ that is proportional to the constraint itself, which vanishes when the
constraint is satisfied.

The impact of the $N$ dependence is more subtle.
However, a simple way to see that it does not spoil the conservation of the
diffeomorphism constraint is to modify the diffeomorphism constraint to
include the term $N_{,i}\, p_N$ that generates diffeomorphisms of $N$ (and
$p_N$), which vanishes when the primary constraints (\ref{primary})
are satisfied
\footnote{Such a modified 
constraint was considered in Ref.~\cite{Kluson2010b} but not explicitly identified as a generator of diffeomorphisms.}.
Similarly one could add the term that generates diffeomorphisms
of $N^i$.
The resulting extended diffeomorphism constraint $\widetilde C_i $ is defined such that
\bea
\int \xi^i \widetilde C_i
&=& \int p_N \mathcal{L}_{\xi} N + p_i \mathcal{L}_\xi N^i + p^{ij}\mathcal{L}_{\xi} g_{ij}\nn \\
&=& \int \xi^i \Big[ (\nabla_i N) p_N + \mathcal{L}_{\vec N} p_i -2 g_{ik} \nabla_j p^{jk} \Big]
%&=& \int \xi^i \Big[ -2 g_{ik} \nabla_j p^{jk} + (\nabla_i N) p_N \nn \\
%&& \quad \qquad + p_j \nabla_i N^j + \nabla_j (p_i N^j) \Big]
\eea
where $\mathcal{L}_{\xi}$ is the Lie derivative.
This extended constraint generates diffeomorphisms of {\it all} the variables, so it actually does have a vanishing bracket with the Hamiltonian.
Since the secondary constraints (\ref{secondary}) already imply that the primary constraints 
(\ref{primary}) are preserved in time, the time independence of $\widetilde{C}_i$ also implies the preservation of the secondary constraint $C_i=0$.

\subsubsection{Propagation of scalar constraint}

Next we turn to the issue of propagation in time of the scalar secondary constraint $C=0$.
First we recall how it works in GR, then we consider the case of nonprojectable Ho\v{r}ava gravity without the $a_i$ dependence in $V$, and finally we include the effects of this dependence.
In GR, the potential $V=-\xi R$ is independent of $N$, 
so that $C=\mathcal{H}_t$.
Moreover, these constraints are first class;
i.e., their Poisson brackets with each other
are combinations of the constraints themselves.
This is obvious for the Poisson brackets with $C_i$,
since $C_i$ just generates diffeomorphisms. The only other bracket
is \cite{Hojman1976}
\beq
\{C(x),C(y)\} = \Big(g^{ij}(x)C_i(x)+g^{ij}(y)C_i(y)\Big)\delta_{,j}(x,y),
\eeq
which closes on the $C_i$ constraint (for any value of $\xi$). 
Thus the time derivative
of $C$ is a combination of the constraints: hence the
constraints are preserved in time.

In Ho\v{r}ava gravity, let us consider first the case where $V(g_{ij})$ is independent of $N$.
Then, when $\lambda\ne1$ and/or $V\ne-\xi R$, the bracket of two $C$'s does not close on a constraint, so the constraints are not first class.
Preservation of $C$ then imposes a further tertiary constraint that depends on $g_{ij}$, $p^{ij}$, and $N$.
This case was analyzed in Ref.~\cite{Henneaux2009} (see also Ref.~\cite{Pons2010}), where it was shown that for generic $g_{ij}$ and $p^{ij}$, the only solution for $N$ is $N = 0$. 
This is unacceptable since the kinetic term of the action \eqref{action} is proportional to $1/N$ and the Hamiltonian generates no evolution, so we will assume $N\ne0$ everywhere. 
When $V = -\xi R$, the tertiary constraint then holds if and only if $\pi/\sqrt{g}$ is constant, and the theory is equivalent to GR in constant mean curvature gauge (provided such a gauge is accessible) \cite{Bellorin2010a}.
The case where $V$ is independent of $N$ has also been considered in the presence of $R^2$-type terms both in the linearized\cite{Das2011} and nonlinear\cite{Bellorin2010b} settings, where it was found that the structure of the constraint algebra as well as the number of propagating degrees of freedom depend on which $R^2$ terms are included.
In what follows we will consider the generic case where $V$ depends on $N$ 
and $\alpha \neq 0$.

In the generic case, $N$ and its spatial derivative occur in
the $a_i$ dependence of $V(g_{ij},a_i)$, so that the constraint $C$
does not commute with $p_N$; i.e., $p_N$ and $C$ are second class.
A possible way to proceed is to simply solve the constraint $C(x) = 0$ 
for one of the dynamical variables, such as $N$ as suggested in 
Ref.~\cite{Kluson2010b}.
The subtlety in doing this reduction is that 
$N$ can only be so determined up to an arbitrary time-dependent,
spatially constant multiple. This is related to the fact that
among the constraints $p_N(x)$ and $C(x)$ are two linear combinations that are first class.
These first-class combinations should not be set to zero strongly, since the symplectic form pulled back to 
such a subspace would be degenerate, and therefore the Poisson brackets would be ill-defined \cite
{Henneaux1994}.
Thus to carry out the reduction we identify the full set of first-class constraints.

\subsection{First-class constraints and Hamiltonian}
For convenience in the analysis, we write the Lagrange multiplier $v$ in the form
\beq
v = N w +N^i \nabla_i N,
\eeq
where the function $w$ is initially arbitrary.
In terms of $w$ the Hamiltonian density \eqref{H} takes the form
\beq\label{HCtilde}
\mathcal{H} = N \mathcal{H}_t + N^i \widetilde C_i + v^i p_i + N w p_N
\eeq
up to a total derivative.
% Note that it is not necessary to perform a similar redefinition of v^i since the Hamiltonian generates diffeomorphisms along N^i and the Lie derivative of N^i along N^i is zero.
The condition that $C$ be preserved in time is then
\beq\label{CHbracket}
\{C(x),H\}=\int N \Bigl( \{C(x), \mathcal{H}_t \}
+ w \{C(x),p_N\} \Bigr) = 0
\eeq
where we have dropped the term proportional to $\{ \widetilde C^i, C \}$, which vanishes when $C = 0$.
By solving for $w$, we ensure that the constraint $C = 0$ is preserved in time, assuming it has been solved at some initial time.
Thus rather than solving the nonlinear equation $C = 0$ for $N$
separately at each instant of time, one can instead solve a linear equation for $w$.
The presence of the free function $w$ can solve the overconstraining problem, provided the 
bracket \eqref{CHbracket} can be set to zero by solving for $w$.

The Lagrange multiplier $w$ is not completely determined by preservation of $C$, instead the 
Eq.~\eqref{CHbracket} determines $w$ up to a solution of the homogeneous equation
\beq \label{homogeneous}
\int N w \{C(x),p_N \} = 0.
\eeq
Each such solution is a gauge freedom in the evolution and corresponds to a primary 
first-class constraint, $\int w N p_N$.
In this case $w(x) = \kappa$ is such a solution, where $\kappa$ is constant.
The existence of this solution follows from the fact that the constraint
\beq \label{npn}
\int N p_N
\eeq
generates constant rescalings of $N$, and $C$ is invariant under such a rescaling.

Equation \eqref{CHbracket} is a linear partial differential equation for $w$ of up to sixth order.
In the IR limit it reduces to a second-order equation,
\begin{equation}
%\nabla^2 w + a^i \nabla_i w = -\frac{1}{2 \alpha \sqrt{g}} \left\{C, \int N \mathcal{H}_t \right\}.
\partial_i( N \sqrt{g} g^{ij} \partial_j w) = -\frac{N}{2 \alpha} \left\{C, \int N \mathcal{H}_t \right\}.
\end{equation}
This equation is elliptic, and in fact the left-hand side can be written as $\sqrt{\tilde g} \tilde \nabla^2 w$ 
where $\tilde \nabla$ is the covariant derivative compatible with the metric $\tilde g_{ij} = N^2 g_{ij}$.
It follows that the solution exists for $w$ provided the integral of the right-hand side is zero,
\beq
\left \{ \int N C, \int N \mathcal{H}_t  \right \} = 0
\eeq
which follows immediately from the definition of $C$ \eqref{C}.
This solution is unique up to the addition of the homogeneous solution $w(x) =\kappa$.

Beyond the IR limit, the equation determining $w$ is 
of higher order and might admit nonconstant homogeneous solutions.
However any theory admitting additional solutions to Eq.~\eqref{homogeneous} will have additional first-class constraints, which may be regarded as generators of gauge transformations \cite{Henneaux1994}.
For a generic choice of parameters, we expect there will be no additional gauge symmetry, so that the general solution of \eqref{CHbracket} has the form
\beq \label{wbar}
w = \overline w[g_{ij},p^{ij},N] + \kappa,
\eeq
where $\kappa$ is constant and $\overline w$ is a particular solution of Eq.~\eqref{CHbracket}.

The solution for $w$ can be substituted back into the Hamiltonian to obtain a Hamiltonian that preserves the constraints in time,
\beq \label{H2}
H = \int N (\mathcal{H}_t + \overline w p_N) + \kappa \int N p_N + \int \left( N^i \widetilde C_i + v^i p_i \right).
\eeq
By varying the Hamiltonian by a constant rescaling of $N$, we can see that the first two terms in Eq.~\eqref{H2} are the Poisson brackets of the Hamiltonian with the first-class constraint $\int N p_N$,
\beq
\left\{ H, \int N p_N, \right\} = \int N (\mathcal{H}_t + \overline w p_N).
\eeq
The Hamiltonian $H$ is therefore a sum of first-class constraints generating the two types of foliation-preserving diffeomorphisms: global time reparametrizations, and spatial diffeomorphisms.

Observables in Ho\v{r}ava gravity must be gauge-invariant functions; 
that is they must commute with all first-class constraints.
Since the Hamiltonian is a sum of first-class
constraints, observables must also have zero 
Poisson bracket with the Hamiltonian and hence be conserved in time.
For example, the volume of the spatial slice $\Sigma$ is diffeomorphism-invariant, but is not time-independent since it fails to commute with the constraint $\int N C$:
\beq
\left \{ \int N C, \int \sqrt{g} \right \} = \frac{1}{1 - 3 \lambda} \int N p.
\eeq
Hence the volume is not an observable in Ho\v{r}ava gravity, contrary to what was claimed in Ref.~\cite{Kluson2010b}.
This reflects the fact that it is meaningless to label an observable by the ``$t$'' coordinate
in a theory that has $t$-reparametrization symmetry.

\section{Reduced phase space} \label{reduce}

We have expressed 
Ho\v{r}ava gravity  as a Hamiltonian system with second-class constraints.
In the presence of second-class constraints it is possible to reduce the phase space by solving the constraints for one or more of the dynamical variables as a function of the others.
The Hamiltonian and symplectic form are then restricted to the reduced phase space, and the restriction of the symplectic form defines a nondegenerate
Poisson bracket on the reduced phase space known as the 
``Dirac bracket'' \cite{Dirac1958, Henneaux1994}.

In Refs.~\cite{Blas2009b,Kluson2010b}, it was proposed that the constraint $C = 0$ \eqref{secondary} be solved for $N$.
There are two related issues with this strategy.
First, $N$ can be determined only up to a time-dependent rescaling.
Second, among the constraints are two first-class constraints; setting these to zero strongly would result in a degenerate symplectic form and therefore an undefined Dirac bracket.

To reduce the system, we need to impose as many linear combinations of the constraints as possible without setting the Hamiltonian to zero.
In order to have a set of constraints that determines $N$ completely, we choose the gauge in which the average lapse is one.
Once this new constraint is introduced the only first-class constraint that remains (apart from the diffeomorphism constraints) is the Hamiltonian.
We therefore impose the following constraints strongly,
\beq \label{constraints}
\int N \sqrt{g} = \int \sqrt{g},\qquad  p_N = 0,\qquad C = C_0 \sqrt{g}
\eeq
and eliminate $N$ from the phase space.
Here $C_0$ is a function of time whose presence is necessary to ensure that the Hamiltonian is not set to zero strongly.
$C_0$ can be expressed in terms of $N$ by integrating the identity $C = C_0 \sqrt{g}$, or alternatively, since $N$ is completely determined by the constraints, $C_0$ can be written in terms of the metric variables $g_{ij}, p^{ij}$.

The reduced phase space has coordinates $g_{ij}, p^{ij}$ and its dynamics are expressed in terms of the Hamiltonian and the Dirac brackets \cite{Henneaux1994}.
Although generically the Dirac brackets are different from the Poisson brackets, in the case where the second-class constraints can be written in the form $N = f[g_{ij}, p^{ij}]$, $p_N = 0$, 
the Dirac and Poisson brackets between $g_{ij}$ and  $p^{ij}$ coincide.
Since the constraints \eqref{constraints} do not restrict the values of $g_{ij}$ and $p^{ij}$, they are of this special form and the Dirac bracket $\{\cdot,\cdot \}^*$ on the reduced phase space is the same as the canonical Poisson bracket,
\beq \label{dirac}
\{ g_{ij}(x), p^{kl}(y) \}^* = \{ g_{ij}(x), p^{kl}(y) \} = \delta^{(k}_i \delta^{l)}_j \delta(x,y).
\eeq
The Hamiltonian on the reduced phase space is found by substituting the second-class constraints \eqref{constraints} into the Hamiltonian \eqref{H2}:
\beq
H = {\cal V} C_0 + \int \left( N^i \widetilde C_i + v^i p_i \right)
\eeq
where $\cal V$ is the volume of the spatial slice $\Sigma$.
Note that the terms involving $\overline w$ are absent, since their only role was to preserve the constraint $C = 0$, which always holds on the reduced phase space.
The equations of motion are then
\bea
\dot{g}_{ij} &=& {\cal V} \frac{\delta C_0}{\delta p^{ij}} + \mathcal{L}_{\vec N} g_{ij} \\
\dot{p}^{ij} &=& - {\cal V} \frac{\delta C_0}{\delta g_{ij}} + \mathcal{L}_{\vec N} p^{ij}
\eea
where we have used the fact that $C_0 = 0$ on shell.
% (Note that on the reduced phase space the diffeomorphism constraint generates diffeomorphisms of $g, N^i$ and their conjugate momenta. The $N$ terms drop out when we set $p_N = 0$.)

In the IR limit, $-C_0$ can be expressed as the smallest eigenvalue of a linear operator,
\beq \label{eigen0}
Ln = -C_0 n,
%\left(4 \alpha \nabla^2 + \frac{1}{g} \left(p^{ij}p_{ij} + \frac{\lambda}{1 - 3 \lambda} p^2\right) - \xi R \right) n
\eeq
where $L$ is given by Eq.~\eqref{L2}, and $n=\sqrt{N}$.
The time evolution then depends on first-order changes of the eigenvalue $C_0$ with respect to $g$, and $p$.
These can be found by first-order perturbation theory, e.g.
\beq
{\cal V} \frac{\delta C_0}{\delta g_{ij}} = -\frac{\delta}{\delta g_{ij}} \int \sqrt{g} \, n Ln
\eeq
and similarly for $p$.
This formula can then be rewritten in terms of $N$:
\bea
{\cal V} \frac{\delta C_0}{\delta p^{ij}} &=& \frac{2 N}{\sqrt{g}} \left( p_{ij} + \frac{\lambda}{1 - 3 \lambda} p g_{ij} \right), \\
{\cal V} \frac{\delta C_0}{\delta g_{ij}} &=&
\sqrt{g} N \left(\alpha a_k a^k + \xi R \right)g^{ij} \nn \\ % vary det g in potential term
&&{} + \frac{N}{\sqrt{g}}\left( p^{kl} p_{kl}+ \frac{\lambda}{1 - 3 \lambda} p^2 \right) g^{ij} \nn \\ % vary det g in kinetic term
&&{} + \alpha \sqrt{g} N a^i a^j \nn \\ %From the laplacian term: just integrate by parts before varying
&&{} + \frac{2 N}{\sqrt{g}} \left(p^{ik}p_k^{\phantom{k} j} + \frac{\lambda}{1 - 3 \lambda} p p^{ij} \right) \nn \\ % vary g's in the index contractions
&&{} + \xi \sqrt{g} \left( \nabla^i \nabla^j - R^{ij} - g^{ij} \nabla^2 \right) N. % Vary the Ricci scalar
\eea
These equations are equivalent to those obtained by varying the Lagrangian \eqref{action}.

\section{Asymptotically flat case} \label{asymflat}

In the preceding discussion, it was assumed that the spatial manifold $\Sigma$ is compact with no boundary.
When there is a boundary (or in the case of asymptotic flatness, an asymptotic region with prescribed falloff conditions), variation of the Hamiltonian will result in a total derivative that can be written as a boundary term.
To have a well-defined variational principle, additional terms must be added to the Hamiltonian to make this boundary term vanish.
The appropriate boundary terms in the asymptotically flat case are determined by the falloff conditions on the fields,
\beq \label{falloff}
g_{ij} = \delta_{ij} + O(r^{-1}), \quad
K_{ij} = O(r^{-2}), \quad
R = O(r^{-3})
\eeq
where $r$ is the radial coordinate of a coordinate system in which the metric asymptotically approaches the Euclidean metric $\delta_{ij}$.
In order for the Hamiltonian to define an asymptotic time translation, we must have $N^i \to 0$ as $r \to \infty$, and we choose a falloff on $N$ such that
\beq \label{falloffN}
N = 1 + O(r^{-1}), \quad \nabla_i N = O(r^{-2})
\eeq
as $r \to \infty$.
The contributions to the boundary term $H_\partial$ in the Hamiltonian then come only from the potential $V$.
For any term in $V$ that is fourth order or higher in derivatives, the corresponding boundary term is of at least third order in derivatives and hence vanishes subject to \eqref{falloff}.
The remaining terms in $V$, which are of second order in derivatives, are $R$ and $a^2$.
The $a^2$ term does not contribute a boundary term because $a_i \delta N = O(r^{-3})$, and the variation of $R$ with respect to $g_{ij}$ can be canceled by adding to the Hamiltonian the boundary term
\bea
H_{\partial} = \xi \oint (\partial_i g_{jk} - \partial_j g_{ik}) \delta^{ik} n^j.
\eea
Here $\oint = \oint_{\partial \Sigma} d^2\sigma \sqrt{h}$ where $\sigma$ are coordinates on the sphere at infinity $\partial \Sigma$, $h$ is the induced metric on $\partial \Sigma$, and $n^i$ is the outward unit normal.
This term is nothing but the usual boundary term for general relativity, the ADM mass, with a factor of $\xi$ coming from the action \eqref{action}.
We have therefore obtained a Hamiltonian that leads to a consistent variational principle in the asymptotically flat setting, given by
\beq
H = \int \mathcal{H} + H_{\partial},
\eeq
where the first term is given by (\ref{H}-\ref{hamiltonian}).

The primary and secondary constraints \eqref{primary} and \eqref{secondary} remain the same in the asymptotically flat setting.
The difference comes when solving for the undetermined part of the Lagrange multiplier $w$ to ensure the propagation of the constraint $C$ \eqref{homogeneous}.
In the asymptotically flat setting there is no global first-class constraint $\int N p_N$, when the global time-reparametrization symmetry is broken by the asymptotic value of $N$.
Formally, the first-class constraint is absent because in order for the flow generated by $\int w N p_N$ to preserve the boundary condition \eqref{falloffN}, we must have $w(x) \to 0$ as $x \to \infty$.
Therefore the homogeneous solution $w = 1$ to Eq.~\eqref{homogeneous} is not admissible and the constraints $p_N$ and $C$ are purely second-class.

The Hamiltonian can be expressed as a sum of bulk constraints \eqref{hamiltonian} and a boundary term by multiplying the definition of $C$ \eqref{C} by $N$ and integrating
\beq \label{boundary}
\int N \mathcal{H}_t = \int N C + \oint n_i V^i.
\eeq
Terms in $V$ with four or more derivatives contribute terms in $V^i$ with three or more derivatives, and these vanish subject to the falloff conditions \eqref{falloff} so that we can replace $V^i$ with its IR limit, Eq.~\eqref{VIR}.
Combining these results with \eqref{HCtilde} we find that the Hamiltonian is
\beq
H = \int \left( N C + \overline w N p_N + N^i \widetilde C_i + v^i p_i \right) 
+ \mathcal{E},
\eeq
which is a sum of local constraints, plus a boundary term $\mathcal{E}$, the total energy.
The total energy $\mathcal{E}$ is the on-shell value of the Hamiltonian, which is the sum of two boundary terms
\beq
\mathcal{E} = \xi \oint (\partial_i g_{jk} - \partial_j g_{ik}) \delta^{ik} n^j
 - 2 \alpha \oint n^i \partial_i N.
\eeq
We emphasize that this is the total energy for the full Ho\v{r}ava gravity,
not just in the IR limit.\footnote{An expression for the total energy in 
Horava gravity was recently found in Ref.~\cite{Blas2011}
using the Noether charge formalism.}

In the limit of large $r$, the metric can be treated as a perturbation of flat spacetime. 
In Newtonian gauge,
\beq
N = 1 + \psi, \qquad g_{ij} = (1 - 2 \phi) \delta_{ij},
\eeq
the energy takes the form
\beq \label{energyphipsi}
\mathcal{E} = \oint n^i \partial_i (4 \xi \phi - 2 \alpha \psi).
\eeq
Note that the ADM energy in GR depends only on $\phi$, whereas in Ho\v{r}ava gravity there is also dependence on $\psi$.

In the static spherically symmetric solutions, $\phi = \psi$ and the asymptotic form of the solution to $O(1/r)$ is \cite{Blas2009b}
\beq \label{asymptotic}
N = 1 - \frac{r_0}{2 r}, \qquad g_{ij} = \left(1 + \frac{r_0}{r} \right) \delta_{ij},
\eeq
where $r_0$ is a parameter with dimensions of length.
Returning to units in which $16 \pi G_H \neq 1$, the energy of this solution is
\beq \label{energy}
\mathcal{E} = \frac{r_0}{2G} \left( \xi - \tfrac{1}{2} \alpha \right).
\eeq
The gravitational constant that appears in the Newtonian limit is not 
$G_H$ but $G_N \equiv G_H / (\xi - \frac{1}{2}\alpha)$ \cite{Blas2009b,Jacobson2010}.
Thus, if we identify $r_0 = 2 G_N M$ in the weak field limit then $\mathcal{E} = M$.

Horava gravity 
in the IR limit 
can be thought of as the hypersurface-orthogonal restriction
of Einstein-aether theory, where the restriction is at the level of the 
Lagrangian\cite{Blas2009a, Germani2009, Jacobson2010}.
Because of this close relation, it is not surprising that the total energy 
expression \eqref{energyphipsi} is identical to the expression for total energy in 
Einstein-aether theory found in Ref.~\cite{Foster2005},
when $\nabla_i \ln N$ is identified with the acceleration of the aether, 
and when the parameters of Ho\v{r}ava gravity are identified 
with the parameters of Einstein-aether theory as in Ref.~\cite{Jacobson2010}.

The reduction of the phase space with asymptotically flat boundary conditions proceeds in much the same way as in the compact case.
The difference is that the constraints $C$ are all second class, so the equation $C = 0$ can be solved strongly as a differential equation for $N$ with the boundary condition $N \to 1$ at spatial infinity.
The reduced phase space has coordinates $g_{ij}, p^{ij}$ with the canonical Poisson brackets \eqref{dirac}.
The Hamiltonian on the reduced phase space is 
\beq
H = \int N C + \int \left(N^i \widetilde C_i + v^i p_i \right) + \mathcal{E},
\eeq
where $N$ is treated as a functional of $g_{ij}$ and $p^{ij}$
defined by the constraint.

\section{Conclusion}

We have derived the Hamiltonian formulation of Ho\v{r}ava gravity for both closed and asymptotically flat spatial manifolds, extending the analysis initiated in Ref.~\cite{Kluson2010b}. 
In contrast with GR, the scalar constraints are not generators of surface deformations.
These constraints are of second class, except for 
a single linear combination that is first class, which 
generates transformations between leaves of the preferred foliation.
The Hamiltonian is a sum of constraints that generate three-dimensional diffeomorphisms and global reparametrizations of time.

We have also considered a phase space reduction of Ho\v{r}ava gravity in which the second-class constraints are solved formally for the lapse.
This is complicated by the fact that the constraints have a global first-class part that cannot be set to zero strongly.
The second-class constraints take the form of a nonlinear differential equation for $N$, but in the IR limit they are linear in $\sqrt{N}$.
In the IR limit, the Hamiltonian is expressible in terms of the least
eigenvalue of a certain differential operator, and is constrained to vanish.

In the asymptotically flat setting, the time-reparametrization symmetry is fixed by the constant value of the lapse at infinity, so the associated global constraint is absent.
Instead the Hamiltonian acquires additional boundary terms that contribute to the total energy.
The energy depends on both Newtonian potentials $\phi$ and $\psi$, not just on $\phi$ as in GR, in exact agreement with the result from Einstein-aether theory.
For the static spherically symmetric solutions, $\phi = \psi$ and the energy takes the same form as GR, with a rescaled Newton
constant.
We do not know whether there are solutions in which $\phi \neq \psi$, for which the energy would not be just a rescaling of the GR energy.\\

{\bf Note added}: Shortly after version one of this paper appeared on the arXiv,
J. Bellor\'in and A. Restuccia posted Ref.~\cite{Bellorin2011}, which treats many of the same issues addressed here. 

\section*{Acknowledgments}
We acknowledge helpful correspondence with M.~Grillakis and S.~Ghosh.
This research was supported in part by the Foundational Questions Institute
(FQXi Grant No. RFP20816), by NSF Grant No.
PHY-0903572, and by a NSERC PGS-D to WD.

\bibliographystyle{utphys}
\bibliography{horava}

\providecommand{\href}[2]{#2}\begingroup\raggedright\begin{thebibliography}{10}

\bibitem{Horava2009}
P.~Ho\v{r}ava, ``{Quantum Gravity at a Lifshitz Point},''
  \href{http://dx.doi.org/10.1103/PhysRevD.79.084008}{{\em Phys. Rev. D}
  {\bfseries 79} (2009) 084008},
\href{http://arxiv.org/abs/0901.3775}{{\ttfamily arXiv:0901.3775 [hep-th]}}.
%%CITATION = 0901.3775;%%.

\bibitem{Sotiriou2010}
T.~P. Sotiriou, ``{Ho\v{r}ava-Lifshitz gravity: a status report},''
  \href{http://dx.doi.org/10.1088/1742-6596/283/1/012034}{{\em J. Phys. Conf.
  Ser.} {\bfseries 283} (2011) 012034},
  \href{http://arxiv.org/abs/1010.3218}{{\ttfamily arXiv:1010.3218 [hep-th]}}.

\bibitem{Blas2009b}
D.~Blas, O.~Pujol\`{a}s, and S.~Sibiryakov, ``{Consistent Extension of
  Ho\v{r}ava Gravity},''
  \href{http://dx.doi.org/10.1103/PhysRevLett.104.181302}{{\em Phys. Rev.
  Lett.} {\bfseries 104} (2010) 181302},
\href{http://arxiv.org/abs/0909.3525}{{\ttfamily arXiv:0909.3525 [hep-th]}}.
%%CITATION = 0909.3525;%%.

\bibitem{Kluson2010b}
J.~Kluson, ``{Note About Hamiltonian Formalism of Healthy Extended
  Ho\v{r}ava-Lifshitz Gravity},''
  \href{http://dx.doi.org/10.1007/JHEP07(2010)038}{{\em JHEP} {\bfseries 07}
  (2010) 038},
\href{http://arxiv.org/abs/1004.3428}{{\ttfamily arXiv:1004.3428 [hep-th]}}.
%%CITATION = 1004.3428;%%.

\bibitem{Isham1992}
C.~J. Isham, ``Canonical Quantum Gravity and the Problem of Time,''
  \href{http://arxiv.org/abs/gr-qc/9210011}{{\ttfamily arXiv:gr-qc/9210011}}.

\bibitem{Jacobson2010}
T.~Jacobson, ``{Extended Ho\v{r}ava gravity and Einstein-aether theory},''
  \href{http://dx.doi.org/10.1103/PhysRevD.81.101502}{{\em Phys. Rev. D}
  {\bfseries 81} no.~10, (May, 2010) 101502};
T.~Jacobson, ``{Erratum: Extended Ho\v{r}ava gravity and Einstein-aether
  theory},'' \href{http://dx.doi.org/10.1103/PhysRevD.82.129901}{{\em Phys.
  Rev. D} {\bfseries 82} no.~12, (December, 2010) 129901(E)},
  \href{http://arxiv.org/abs/1001.4823}{{\ttfamily arXiv:1001.4823 [hep-th]}}.

\bibitem{Carruthers2010}
I.~Carruthers and T.~Jacobson, ``{Cosmic alignment of the aether},''
  \href{http://dx.doi.org/10.1103/PhysRevD.83.024034}{{\em Phys. Rev. D}
  {\bfseries 83} (2011) 024034},
  \href{http://arxiv.org/abs/1011.6466}{{\ttfamily arXiv:1011.6466 [gr-qc]}}.

\bibitem{Blas2009a}
D.~Blas, O.~Pujolas, and S.~Sibiryakov, ``{On the Extra Mode and Inconsistency
  of Ho\v{r}ava Gravity},''
  \href{http://dx.doi.org/10.1088/1126-6708/2009/10/029}{{\em JHEP} {\bfseries
  10} (2009) 029},
\href{http://arxiv.org/abs/0906.3046}{{\ttfamily arXiv:0906.3046 [hep-th]}}.
%%CITATION = 0906.3046;%%.

\bibitem{Barausse2011}
E.~Barausse, T.~Jacobson, and T.~P. Sotiriou, ``{Black holes in Einstein-aether
  and Ho\v{r}ava-Lifshitz gravity},''
  \href{http://dx.doi.org/10.1103/PhysRevD.83.124043}{{\em Phys. Rev. D}
  {\bfseries 83} (2011) 124043},
  \href{http://arxiv.org/abs/1104.2889}{{\ttfamily arXiv:1104.2889 [gr-qc]}}.

\bibitem{Courant1989}
R.~Courant and D.~Hilbert, {\em {Methods of mathematical physics}}.
\newblock Wiley-Interscience, 1989.

\bibitem{Henneaux1994}
M.~Henneaux and C.~Teitelboim, {\em {Quantization of Gauge Systems}}.
\newblock Princeton University Press, August, 1994.

\bibitem{Hojman1976}
S.~A. {Hojman}, K.~{Kucha{\v r}}, and C.~{Teitelboim}, ``{Geometrodynamics
  regained},'' \href{http://dx.doi.org/10.1016/0003-4916(76)90112-3}{{\em
  Annals of Physics} {\bfseries 96} (Jan., 1976) 88--135}.

\bibitem{Henneaux2009}
M.~Henneaux, A.~Kleinschmidt, and G.~L. Gomez, ``{A dynamical inconsistency of
  Ho\v{r}ava gravity},''
  \href{http://dx.doi.org/10.1103/PhysRevD.81.064002}{{\em Phys. Rev. D}
  {\bfseries 81} (2010) 064002},
  \href{http://arxiv.org/abs/0912.0399}{{\ttfamily arXiv:0912.0399 [hep-th]}}.

\bibitem{Pons2010}
J.~M. Pons and P.~Talavera, ``{Remarks on the consistency of minimal deviations
  from General Relativity},''
  \href{http://dx.doi.org/10.1103/PhysRevD.82.044011}{{\em Phys. Rev. D}
  {\bfseries 82} (2010) 044011},
  \href{http://arxiv.org/abs/1003.3811}{{\ttfamily arXiv:1003.3811 [gr-qc]}}.

\bibitem{Bellorin2010a}
J.~Bellor\'in and A.~Restuccia, ``{On the consistency of the Ho\v{r}ava
  Theory},''
\href{http://arxiv.org/abs/1004.0055}{{\ttfamily arXiv:1004.0055 [hep-th]}}.
%%CITATION = 1004.0055;%%.

\bibitem{Das2011}
S.~Das and S.~Ghosh, ``{Gauge Invariant Extension of Linearized Ho\v{r}ava
  Gravity},'' \href{http://arxiv.org/abs/1104.1975}{{\ttfamily arXiv:1104.1975
  [gr-qc]}}.

\bibitem{Bellorin2010b}
J.~Bellor\'in and A.~Restuccia, ``{Closure of the algebra of constraints for a
  non-projectable Ho\v{r}ava model},''
  \href{http://dx.doi.org/10.1103/PhysRevD.83.044003}{{\em Phys. Rev. D}
  {\bfseries 83} (2011) 044003},
  \href{http://arxiv.org/abs/1010.5531}{{\ttfamily arXiv:1010.5531 [hep-th]}}.

\bibitem{Dirac1958}
P.~A.~M. Dirac, ``{Generalized Hamiltonian Dynamics},''
  \href{http://dx.doi.org/10.1098/rspa.1958.0141}{{\em Proc. R. Soc. Lond. A}
  {\bfseries 246} no.~1246, (1958) 326--332}.

\bibitem{Blas2011}
D.~Blas and H.~Sanctuary, ``{Gravitational Radiation in Ho\v{r}ava Gravity},''
  \href{http://dx.doi.org/10.1103/PhysRevD.84.064004}{{\em Phys. Rev. D}
  {\bfseries 84} (Sep, 2011) 064004},
  \href{http://arxiv.org/abs/1105.5149}{{\ttfamily arXiv:1105.5149 [gr-qc]}}.

\bibitem{Germani2009}
C.~Germani, A.~Kehagias, and K.~Sfetsos, ``{Relativistic Quantum Gravity at a
  Lifshitz Point},''
  \href{http://dx.doi.org/10.1088/1126-6708/2009/09/060}{{\em JHEP} {\bfseries
  09} (2009) 060},
\href{http://arxiv.org/abs/0906.1201}{{\ttfamily arXiv:0906.1201 [hep-th]}}.
%%CITATION = 0906.1201;%%.

\bibitem{Foster2005}
B.~Z. Foster, ``{Noether charges and black hole mechanics in Einstein-aether
  theory},'' \href{http://dx.doi.org/10.1103/PhysRevD.73.024005}{{\em Phys.
  Rev. D} {\bfseries 73} (2006) 024005},
\href{http://arxiv.org/abs/gr-qc/0509121}{{\ttfamily arXiv:gr-qc/0509121}}.
%%CITATION = GR-QC/0509121;%%.

\bibitem{Bellorin2011}
J.~Bellor\'in and A.~Restuccia, ``{Consistency of the Hamiltonian formulation
  of the lowest-order effective action of the complete Ho\v{r}ava theory},''
  \href{http://arxiv.org/abs/1106.5766}{{\ttfamily arXiv:1106.5766 [hep-th]}}.

\end{thebibliography}\endgroup

\end{document}